**Probing the local structural order of $C_{60}$ thin films by their exciton transport characteristics**

A. K. Topczak[a,], M. Gruber[b], W. Brütting[b] and J. Pflaum[a,c]

[a]Julius-Maximillians University, Am Hubland, 97074 Würzburg, Germany

[b]Institut für Physik, University of Augsburg, 86135 Augsburg

[c]ZAE Bayern, Am Hubland, 97074 Würzburg, Germany

Abstract:

We investigate effects of the local structure on the excitonic transport in $C_{60}$ thin films by means of temperature-dependent photoluminescence quenching studies in a range of 5 to 300 K. Exciton motion in the X-ray amorphous layers indicates thermally activated transport initiated above 80 K accompanied by a thermalization of excitations into low lying states within the excitonic density of states. Discontinuities in the temperature-dependent photoluminescence behavior can be attributed to a continuous phase transition at 80 K and a first order phase transition at 180 K, thereby revealing structural information on the local film morphology in the $C_{60}$ films.

Small exciton diffusion lengths ($L_D$) and low charge carrier mobilities limit the efficiency and performance of organic photovoltaic devices. Previous studies have shown that both transport characteristics strongly depend on the morphology and could be significantly enhanced in polycrystalline layers and even more in organic single crystals[1]. Whereas in inorganic semiconductors valence and conduction bands are several eV in width[2], the weak Van-der-Waals interaction in molecular crystals leads to narrow bandwidths of about 0.5 eV[3,4]. Therefore, the competition between small dispersion and strong phonon coupling directly questions the nature of exciton transport in organic semiconductors, i.e. whether it is related to incoherent motion[5], expected for amorphous, disordered organic layers[3], or to coherent motion[6], which is predominant in organic single crystals and crystalline films at low temperatures[7,8]. As a pivotal criterion for either of the two transport scenarios the occurrence of an effective activation barrier can be utilized. In our study we focus on the exciton transport in $C_{60}$ thin films which are characterized by a Gaussian distribution of energy states due to local energetic disorder as confirmed, e.g. by the incoherent, thermally activated transport of charge carriers[3,5]. The choice of material is motivated by the fact, that $C_{60}$ and its derivatives define model systems for the opto-electronic properties of polyaromatic hydrocarbons with (almost) spherical symmetry. They constitute the most prominent acceptor components organic photovoltaic devices[9]. Finally, the spherical symmetry of the transition dipole orientation in $C_{60}$ renders excellent optical properties, indicated *inter alia* by isotropic absorption coefficient of up to $10^5$ cm$^{-1}$ in the UV-VIS[10]. As will be demonstrated, comparative photoluminescence (PL) quenching studies on $C_{60}$ layers as a function of temperature provide a powerful approach not only to judge on the nature of exciton motion and to quantify the related activation energy[11], but also to probe structural changes on molecular length scales.

To aim for well-defined samples with reliable opto-electronic characteristics thin films of purified $C_{60}$ were vacuum sublimed on commercial glass substrates in a stepped wedge structure. Afterwards, the films were half-side covered with a 10 nm thick exciton blocking layer of Bathophenanthroline (BPhen) to prevent exciton quenching by the top-deposited Ag layer (60 nm), whereas quenching occurs for the $C_{60}$ area without intermediate BPhen layer (see inset Fig. 1a). We chose silver as quenching material to ensure an almost 100% quenching efficiency and to enhance the otherwise weak PL-signal of $C_{60}$ by prolongation of the optical path length and by interference effects of the incident light. The PL-quenching was measured by illuminating both sample areas simultaneously via a cw-Nd:YAG laser with the excitation wavelength of 532 nm being sufficient in energy to excite the relevant optical transitions of $C_{60}$. The PL intensities were spectrally resolved with a CCD camera in combination with a spectrometer. Temperature dependent PL-quenching studies were accomplished between 5 to 300 K. Complementary structural analysis at room temperature renders all $C_{60}$ thin film samples X-ray amorphous which we expect to persist towards lower temperatures by the lack of thermal energy.

Determining the PL intensities of the $C_{60}$ areas with and without intermediated BPhen layer[12], referred to as $PL^{nQ}$ and $PL^Q$ in the following, the quenching ratio $Q = PL^Q / PL^{nQ}$ can be estimated and provides detailed information on the presence of non-radiative decay channels, e.g. by exciton annihilation. With increasing layer thickness, excitons become more and more hindered to reach the quencher interface within their radiative lifetime due to their finite diffusion length $L_D$. As a result, the photoluminescence intensity of the free and the quencher-covered areas will gradually equal each other. To cope with real sample structures, especially upon employing metallic quencher such as Ag, interference effects will modulate the PL-signal significantly[10,13]. Accordingly, we considered the thickness dependent intensity profile and therefore, the local exciton density under continuous illumination, i.e. at steady state conditions. The exciton generation $G(z,t)$ is proportional to the spatial intensity distribution of the incident light. Accounting for light reflection at the silver interface yields a depth dependent oscillation of the intensity profile[13,14]. Hence the one-dimensional exciton diffusion equation at steady state

$$\frac{dn(z,t)}{dt} = D\frac{\partial^2 n(z,t)}{\partial z^2} - \frac{n(z,t)}{\tau} + G(z,t) = 0, \qquad (1)$$

can be solved according to the respective boundary conditions at the various interfaces. Suited boundary conditions can be rationalized by neglecting quenching at the glass/organic-interface, $\left(\frac{\partial n(z,t)}{\partial z}\Big|_{z=0} = 0\right)$, as well as at the organic/BPhen interface and by assuming a quenching probability described by the quality factor V at the organic/quencher interface. V considers the situation of an incomplete exciton quenching and is defined by the slope of the exciton concentration gradient at the quencher interface (z=d) for real samples versus that in case of an idealized perfect quenching[15]. Since silver is used as quencher in our samples, surface plasmons will have a strong impact on the PL modulation for both areas of the organic film at a given thickness. The reason is that an excited molecule can not only dissipate its excitation energy by exciton diffusion to the metal interface but also by radiative coupling to a surface plasmon polariton (SPP) mode at the metal-organic interface[16]. While the former process can be suppressed by a 10nm thin BPhen layer, the near-field coupling to SPPs will persist to distances of the order of several ten to 100nm[17]. Thus numerical simulations[18,19] have been employed to correct the measured PL intensities of the $C_{60}$ samples with and without BPhen spacer for the relative contributions of non-emissive coupling of excitons to SPPs (and other trapped optical modes) at all layer thicknesses. We interpret our temperature dependent PL measurements and, hence exciton motion, in a qualitative similar way that has been suggested for single charge carriers in organic semiconductors. Therefore, transport is assigned to either coherent motion, yielding an increase of mobility towards lower temperatures, or to incoherent motion, characterized by an increase of mobility with temperature.

In the following, we assumed effects by triplet excitons to play a negligible role, since their wavefunction is strongly localized and their motion is limited by a short-range Dexter transport. At first we focus on the room temperature PL-quenching as a function of layer thickness, which yields the exciton diffusion length $L_D$ in $C_{60}$ for a temperature range being of relevance for many organic thin film applications, such as photovoltaic cells or light emitting diodes. Representatively, Fig. 1a) illustrates the PL signal of a 50 nm thick sample at room temperature. The main transitions peaks at 680 nm, 730 nm and 770 nm can be associated with surface-related excitonic states, the $S_1$ bulk emission, and low-energetic X traps caused by $C_{60}$ molecules in proximity to chemical impurities, respectively[20]. Moreover, the pronounced decrease in intensity for the quencher covered area becomes obvious. Fig. 1b) displays the relative PL quenching, Q, as a function of $C_{60}$ layer thickness. Explicitly accounting for interference effects in the exciton diffusion model produces the fit shown in Fig. 1b). The deviation for thick $C_{60}$ layers (> 100 nm) results from the assumption of a perfectly smooth Ag quencher interface, i.e. idealized interference pattern, in the surface plasmon corrections. Our fit is based on a diffusion length of $L_D$ = 5 nm at room temperature. This value is slightly smaller in comparison with literature data of 7.7 nm[21], and 40 nm[22], determined by the external quantum efficiency (EQE) of an operational OPV device. The main differences result from superimposed electrical fields, the interaction of excitons with free and trapped charges[23] as well as morphological influences by the complementary donor phase. Together with the usually neglected effect of a reduced exciton lifetime caused by coupling to surface plasmons this might lead to deviations and overestimations of the effective exciton transport layer thickness and therefore, of the related diffusion length. Our corrected data reveal, that the PL signal for small layer thicknesses < 75 nm is strongly affected by the coupling to surface plasmons.

The diffusion length of only 5 nm points out, that structural disorder in our $C_{60}$ films strongly constrains the delocalization of the excitonic wavefunction and thus, the exciton transport, which will be addressed in more detail in the following by temperature dependent PL quenching studies. In case of a coherent (band-like) exciton transport, the relative quenching Q should decrease towards lower temperatures since reduced exciton-phonon scattering results in an overall enhancement of the exciton lifetime τ and therefore of the diffusion length $L_D$. This phenomenon has been evidenced not only in transport studies on free charge carriers but also for excitons in previous studies on long-range ordered films of the molecular semiconductor Diindenoperylene which exhibits a non-activated exciton transport below 80 K[15]. The Q(T) behavior we observed for our $C_{60}$ layers shows the opposite trend. The relative quenching Q normalized to its room temperature value raises with decreasing temperature (see Fig. 2a) according to a thermally activated exciton motion. Consequently, $L_D$ decreases with decreasing temperature and *vice versa* the exciton localization is strongly enhanced at lower temperatures. This case resembles that of conjugated polymers as reported by Mikhnenko and co-workers[24] and can be attributed to an incoherent exciton motion in disordered films of $C_{60}$. Moreover, at 80 K and 180 K two distinct kinks can be identified in the relative quenching data (see Fig. 2a), which hint at the correlation between local film morphology and exciton transport behavior. Both kinks can be assigned to phase transitions reported for $C_{60}$ bulk single crystals[25]. In $C_{60}$ single crystals a continuous second order phase transition takes place at around 90 K, which is characterized by a freeze-out of rotational motion accompanied by a latching of the molecular entities into an energetically favored configuration sparing a small amount of static disorder. This rotational optimization is characterized by a locking of the electron-rich regions of the inter-pentagon bonds to the electron-depleted regions of the pentagons.

In addition, a first order phase transition is reported at 260 K, which refers to a transition from a high symmetry fcc (face centered cubic) crystal structure at higher temperatures to an sc (simple cubic) structure. For both crystal structures, the individual $C_{60}$ molecules are supposed to rotate freely and independently. The fact, that these phase transitions can be observed in our $C_{60}$ thin films despite their X-ray amorphous structure demonstrates the sensitivity of PL measurements to the local surroundings of the individual molecules. Moreover, the shift of the phase transition to lower temperatures with respect to bulk single crystals can be understood by the enhanced surface-to-volume ratio in our thin film samples. According to the Lindemann criterion the phase transition temperature of a solid state scales with the motional degree of freedom of the atomic or molecular constituents[26,27]. Hence, in case of a pronounced surface contribution, e.g. due to the thin film texture, the overall next-neighbor interaction will be strongly reduced and so does the transition temperature. The slightly higher transition temperatures in our $C_{60}$ layers compared to an idealized Lindemann-type behavior, which relates the transition temperatures in 2D and 3D via $T_{2D} \approx T_{3D}^{2/3}$, are owned to deviations of our real samples from a mere 2D structure, e.g. by the presence of the underlying substrate. According to the dimensionality reasons, the temperature dependent PL behavior of the surface-related (2D) excitons at 680nm is expected to differ from that of the bulk (3D) excitons at 730nm. This prospect is evidenced by the corresponding PL spectra, which besides a thermal variation of the relative peak intensities, indicate a distinct blue shift of the main transition peak at 730 nm for temperatures above 180 K (s. Fig. 2b). This blue shift of the steady state PL spectrum with increasing temperature can be explained by the energetic disorder in our samples which causes a broadening of the excitonic density of states (DOS), as previously reported[28,29]. Upon light absorption, excitons are generated in the high energy regime of the excitonic DOS and subsequently thermalize into lower-lying energy states, as illustrated in Fig. 3a). The energetic downward motion ceases when excitons have reached the most likely occupied state of the excitonic DOS. With increasing thermal energy, i.e. temperature, this level continuously shifts towards the center of the excitonic DOS, which causes the observed blue shift in the PL. The phonon-mediated lifting of excitons to higher energy states in the excitonic DOS promotes exciton hopping, as confirmed by our quenching results. Whereas the temperature dependent wavelength shift is mainly assigned to thermalization of the photoexcited states, the kinks in the temperature dependent PL behavior of the bulk exciton (730 nm) and the surface exciton (680 nm) can be attributed to the first order phase transition in $C_{60}$. In agreement with single crystal data[30], the transition appears in the bulk exciton emission at 260 K, followed by a continuous increase of its wavelength down to 160 K (Fig. 2b).

The lower temperature value corresponds to the first order phase transition temperature we deduced from the relative quenching data, based in the integrated PL intensity (Fig. 2a). In addition, the energy difference of the related emission wavelength shift of the volume exciton results in 22 meV, which is in almost perfect agreement with the thermal energy $k_BT$ at 260 K. This result indicates that the bulk exciton DOS is mainly conserved upon phase transition and confirms the energetic relaxation within the excitonic DOS Fig. 3a). By contrast, due to the reduced dimensionality the first order phase transition temperature for the surface exciton is shifted to lower temperatures of 120 K. In this case, the energy difference of corresponding wavelength shift yields 19 meV which in comparison to the thermal energy of 10 meV at 120 K is enhanced by a factor of two. This difference might be attributed to a pronounced change of the surface exciton DOS for the two crystal structures as well as to a spatially anisotropic distribution of thermal energy, caused by the symmetry breaking and the lack of next nearest neighbors at the film surface. The thermally activated transport is initiated above 80 K. Below this temperature the PL intensity remains almost constant, due to a transport behavior governed by fast thermalization of excitons to low energy states of the excitonic DOS, as illustrated in Fig. 3a), disabling a pronounced, phonon-assisted hopping transport by the lack of thermal energy and the number of accessible states[24]. Modeling the logarithm of the PL intensity over $1/k_BT$ by an Arrhenius Plot (See Fig. 3b) we estimated an effective activation energy for the incoherent exciton transport in the range of 25 ± 15 meV. We refer this activation energy to a "self-trapping barrier", which is caused by molecular relaxation and which excitons must overcome to attain adjacent transport states. This energy is reported to be of the order of 12 meV for $C_{60}$ bulk single crystals[30]. The higher activation energy in our $C_{60}$ thin film samples originates from the enhanced structural disorder as well as from boundaries between short-range ordered regions within the organic semiconducting layers. As a consequence, at higher temperatures the effective exciton transport level lies deep in the excitonic DOS and is no longer restricted by thermal activation, which is in agreement with the weak temperature dependence of Q observed above 200 K.

Our photoluminescence (PL) studies prove exciton motion in X-ray amorphous $C_{60}$ films to be governed by an incoherent hopping transport. The related activation energies of the order of 25±15 meV for temperatures above 80 K resemble the "self-trapping barrier" energy determined for $C_{60}$ bulk samples. As demonstrated, both phase transitions reported for single crystals could be identified in the temperature dependent PL data of $C_{60}$ thin films but with the additional advantage of being sensitive to volume and surface related contributions to the excitonic DOS. In summary, the presented approach highlights the possibility of probing structural properties on molecular length scales by excitonic processes in organic semiconductors.

Acknowledgments

Financial support by the DFG focus program SPP1355 is acknowledged. We gratefully appreciate financial support by the Bavarian State Ministry of Science, Research, and the Arts within the Collaborative Research Network ''Solar Technologies go Hybrid''

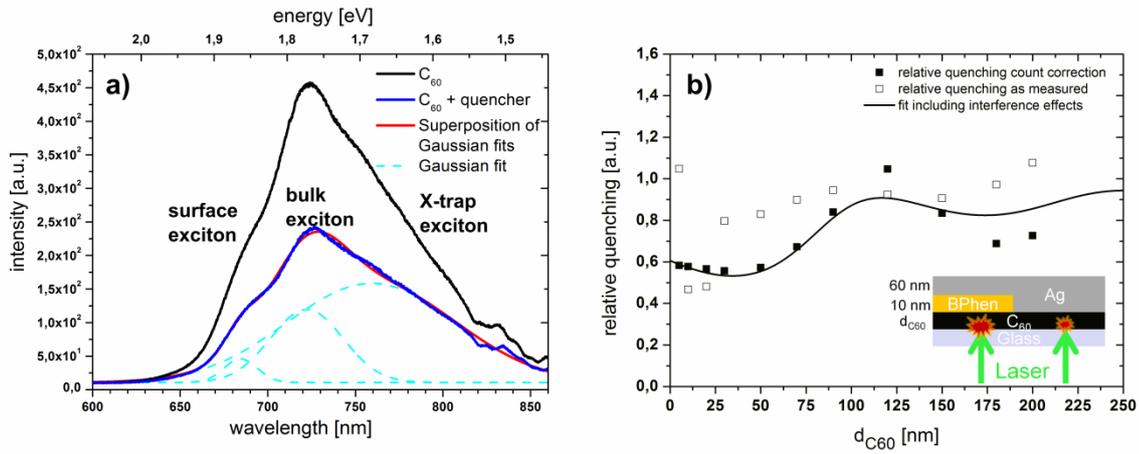

Fig 1: a) PL spectra of a 50 nm thick $C_{60}$ with and without Ag quencher, including Gaussian fits of the three main PL peaks, related to surface excitons (680 nm), the bulk excitation (730 nm) and X traps (770 nm). b) Thickness dependence of the relative quenching, Q, at room temperature, the pristine relative quenching (open dots) and considering corrections due to surface plasmon effects on the exciton lifetime (black dots), Inset: sample geometry.

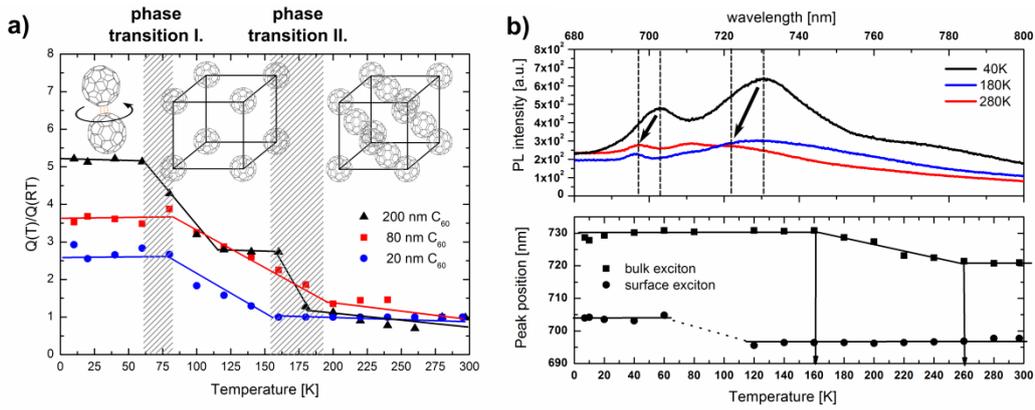

Fig. 2: a) Temperature behavior of the relative quenching Q normalized to its room temperature value Q(RT). Three different regions of the Q(T) behavior can be identified and have been related to different structural phases of $C_{60}$ illustrated in the graph. b) PL spectra of a 200 nm thick $C_{60}$ layer at various temperatures. A pronounced blue shift of the bulk exciton at 730 nm occurs between 160 and 260 K, whereas the corresponding shift of the surface exciton appears already between 60 and 120 K.

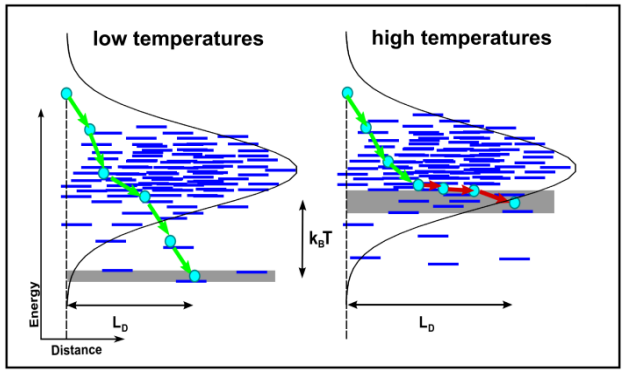

Fig 3: Scheme of the excitonic density of states (DOS) at low temperatures, with the populated state located at the lower tail, and at higher temperatures, with the populated state located almost in the center of the DOS.

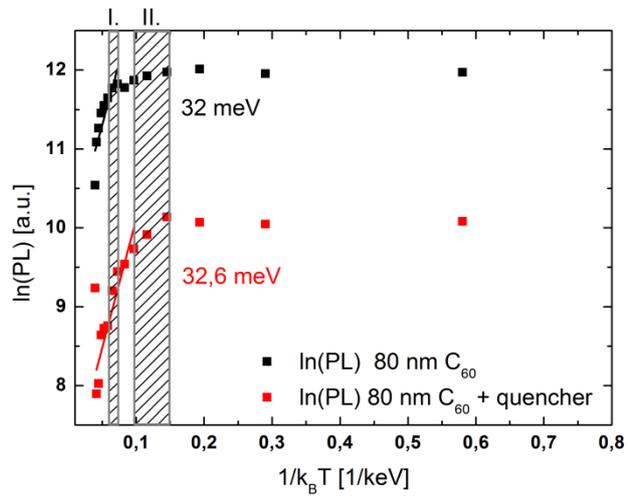

Fig. 4: Arrhenius plots modeling the measured PL intensity of a representative 80 nm thick $C_{60}$ layer with and without quencher (red and black dots, respectively).